\documentclass[twocolumn,10pt,journal,a4paper]{IEEEtran}
\usepackage[latin1]{inputenc}
\usepackage{amssymb}
\usepackage{latexsym}
\usepackage{mathrsfs}
\usepackage{amsfonts}
\usepackage{amsbsy}
\usepackage{amsmath}
\usepackage{times}
\usepackage{epsfig,epsf,times,amssymb,amsmath}
\usepackage{geometry}
\input epsf
\title{\Huge Linear Reduced-Rank Interference Suppression for DS-UWB
Systems Using Switched Approximations of Adaptive Basis Functions}
\author{Sheng Li, Rodrigo C. de Lamare and Rui Fa \vspace{-2em}
\thanks{Copyright (c) 2010 IEEE.
Personal use of this material is permitted. However, permission to use this
material for any other purposes must be obtained from the IEEE by sending a
request to pubs-permissions@ieee.org.

This work is supported by the Department of Electronics, University of York.
The authors are with the Communications Research Group, Department
of Electronics, University of York, York, YO10 5DD, UK (e-mail:
\{sl546,rcdl500 and rf533\}@ohm.york.ac.uk). }}
\begin{document}
\maketitle

\begin{abstract}
In this work, we propose a novel low-complexity reduced-rank scheme and
consider its application to linear interference suppression in direct-sequence
ultra-wideband (DS-UWB) systems. Firstly, we investigate a generic reduced-rank
scheme that jointly optimizes a projection vector and a reduced-rank filter by
using the minimum mean-squared error (MMSE) criterion. Then a low-complexity
scheme, denoted switched approximation of adaptive basis functions (SAABF), is
proposed. The SAABF scheme is an extension of the generic scheme, in which the
complexity reduction is achieved by using a multi-branch framework to simplify
the structure of the projection vector.
Adaptive implementations for the SAABF scheme are developed by using
least-mean squares (LMS) and recursive least-squares (RLS)
algorithms. We also develop algorithms for selecting the branch
number and the model order of the SAABF scheme. Simulations show
that in the scenarios with severe inter-symbol interference (ISI)
and multiple access interference (MAI), the proposed SAABF scheme
has fast convergence and remarkable interference suppression
performance
with low complexity.
\end{abstract}


\section{Introduction}

Ultra-wideband (UWB) technology \cite{rascholtz1993}-\cite{RCQiu2005} is a promising short-range wireless
communication technique with the potential to achieve high data rates. 
In $2005$, direct-sequence ultra-wideband (DS-UWB)
\cite{oppermann2004}-\cite{RFisher2005} was proposed as a possible
standard physical layer technology for wireless personal area
networks (WPANs) \cite{RFisher2005}. For DS-UWB systems, the huge
transmission bandwidths introduce a high degree of diversity at the
receiver due to a large number of resolvable multipath components
(MPCs) \cite{dajana2007}. In multiuser scenarios, the receiver is
required to mitigate the multiple-access interference (MAI) and the
inter-symbol interference (ISI) effectively with affordable
complexity. Possible solutions of interference suppression include the linear schemes and nonlinear schemes such as
the successive interference cancelation (SIC) \cite{Rodrigo20071}
and decision feedback (DF) \cite{Rodrigo20081} schemes.
For DS-UWB communications, the major challenge for the interference
suppression schemes is to obtain fast convergence and satisfactory
steady state performance in dense-multipath environments. 
In conventional full-rank adaptive algorithms, a long filter length is required
for DS-UWB systems and hence, these algorithms confront slow convergence and
poor tracking performance.


To overcome the drawbacks of the full-rank algorithms in UWB
communications, reduced-rank schemes have been recently considered.
A reduced-order finger selection linear MMSE receiver
with RAKE-based structures have been proposed in \cite{SB2005},
which requires the knowledge of the channel and the noise variance.
Solutions for reduced-rank channel estimation and synchronization in
single-user UWB systems have been proposed in \cite{jianzhang2004}.
For multiuser detection in UWB communications, reduced-rank schemes
have been developed in \cite{shwu2004}-\cite{yTian2006} requiring
knowledge of the multipath channel. The reduced-rank filtering techniques have faster convergence than
the full-rank algorithms \cite{GKutz2007}-\cite{Rodrigo2007} and the well-known
reduced-rank techniques include the eigen-decomposition methods such as the
principal components (PC) \cite{AMHaimovich1991} and the cross-spectral metric
(CSM) \cite{JSGoldstein1997}, the Krylov subspace methods such as the powers of
R (POR) \cite{serguei2002}, the multistage Wiener filter (MSWF)
\cite{MLHonig2002},\cite{JSGoldstein1998} and the auxiliary vector filtering
(AVF) \cite{dapados1999}. Eigen-decomposition methods are based on the eigen-decomposition of the
estimated covariance matrix of the input signal. These methods have very high computational complexity and the performance is
Compared with the full-rank linear filtering techniques, the MSWF
and AVF methods have faster convergence speed with a much smaller
filter size. However, their computational complexity is still very
high.

In this paper, we firstly investigate a generic reduced-rank scheme with joint
and iterative optimization of a projection vector and a reduced-rank linear
estimator to minimize the mean square error (MSE) cost function. Since
information is exchanged between the projection matrix and the reduced-rank
filter for each adaptation, this generic scheme outperforms other existing
reduced-rank schemes. However, in this generic scheme, a large projection
vector is required to be updated for each time instant and hence introduces
high complexity. In order to obtain a low-complexity configuration of the
generic scheme and maintain the performance, we propose the novel switched
approximation of adaptive basis functions (SAABF) scheme.
The basic idea of the SAABF scheme is to simplify the design of the projection
vector by using a multiple-branch framework such that the number of
coefficients to be adapted in the projection vector is reduced and hence
achieve the complexity reduction.
The LMS and RLS adaptive algorithms are then developed for the joint adaptation
of the shortened projection vector and the reduced-rank filter. We also propose
adaptive algorithms for branch number selection and model order adaptation. 

The main contributions of this work are listed below.

\begin{itemize}
\item {A novel low-complexity reduced-rank scheme is proposed for interference suppression in DS-UWB system.}
\end{itemize}

\begin{itemize}
\item {LMS and RLS adaptive algorithms are developed for the proposed scheme.}
\end{itemize}

\begin{itemize}
\item {Algorithms for selecting the scheme parameters are proposed.}
\end{itemize}

\begin{itemize}
\item {The relationships between the proposed SAABF scheme, the generic scheme and the full-rank scheme are established.}
\end{itemize}

\begin{itemize}
\item {Simulations are performed with the IEEE 802.15.4a channel model and severe ISI and MAI are assumed for the evaluation of the proposed scheme.}
\end{itemize}

The rest of this paper is structured as follows. Section
\ref{sec:uwbsystem} presents the DS-UWB system model. In Section
\ref{sec:rrscheme}, the design of the generic reduced-rank scheme is
detailed. The proposed SAABF scheme is described in Section
\ref{sec:proposedscheme} and the adaptive algorithms and the
complexity analysis are presented in Section
\ref{sec:adaptivealgorithms}. The proposed adaptive algorithms for
selecting the key parameters of the SAABF scheme are described in
Section \ref{sec:parameters}. Simulations results are shown in
Section \ref{sec:simulations} and conclusions are drawn in Section
\ref{sec:conclusion}.


\section{DS-UWB System Model}
\label{sec:uwbsystem}

In this work, we consider the uplink of a binary phase-shift keying (BPSK)
DS-UWB system with $K$ users. A random spreading code $\mathbf {s}_{k}$ is
assigned to the $k$-th user. The spreading gain is $N_{c} = T_{s}/T_{c}$, where
$T_{s}$ and $T_{c}$ denote the symbol duration and chip duration, respectively.
The transmit signal of the $k$-th user, $k=1,2,\dots,K$, can be expressed as
\begin{equation}
x^{(k)}(t)=\sqrt{E_{k}}\sum_{i=-{\infty}}^{\infty}\sum_{j=0}^{N_{c}-1}
p_{t}(t-iT_{s}-jT_{c})s_{k}(j)b_{k}(i),
\end{equation}where $b_{k}(i)$ $\in$ $\{\pm1\}$
denotes the BPSK symbol for the $k$-th user at the $i$-th time
instant, $s_{k}(j)$ denotes the $j$-th chip of the spreading code
$\mathbf {s}_{k}$. $E_{k}$ denotes the transmission energy.
$p_{t}(t)$ is the pulse waveform of width $T_{c}$. For UWB
communications, widely used pulse shapes include the Gaussian
waveforms, raised-cosine pulse shaping and root-raised cosine (RRC)
pulse shaping \cite{KazimierzSiwiak2004},\cite{RFisher2005}.
Throughout this paper, the pulse waveform $p_{t}(t)$ is modeled as
the RRC pulse with a roll-off factor of $0.5$
\cite{Aparihar2007},\cite{RFisher2005} and \cite{AP2005}.

The channel model considered is the IEEE 802.15.4a standard channel
model for the indoor residential non-line of sight (NLOS)
environment \cite{Molisch2005}. This standard channel model includes
some generalizations of the Saleh-Valenzuela model and takes the
frequency dependence of the path gain into account
\cite{Molisch2006}. In addition, the 15.4a channel model is valid
for both low-data-rate and high-data-rate UWB systems
\cite{Molisch2006}. For the $k$-th user, the channel impulse
response (CIR) of the standard channel model is $h_{k}(t)=\sum_{u=0}^{L_{c}-1}\sum_{v=0}^{L_{r}-1}\alpha_{u,v}e^{j\phi_{u,v}}\delta(t-T_{u}-T_{u,v}),$
where $L_{c}$ denotes the number of clusters, $L_{r}$ is the number
of multipath components (MPCs) in one cluster. $\alpha_{u,v}$ is the
fading gain of the $v$-th MPC in the $u$-th cluster, $\phi_{u,v}$ is
uniformly distributed in $[0,2\pi)$. $T_{u}$ is the arrival time of
the $u$-th cluster and $T_{u,v}$ denotes the arrival time of the
$v$-th MPC in the $u$-th cluster. For the sake of simplicity, we
express the CIR as
\begin{equation}
h_{k}(t)= \sum_{l=0}^{L-1}h_{k,l}\delta(t-lT_{\tau}),\label{eq:channel}
\end{equation} where $h_{k,l}$
and $lT_{\tau}$ present the complex-valued fading factor and the arrival time
of the $l$-th MPC ($l=uL_{c}+v$), respectively. $L=T_{DS}/T_{\tau}$ denotes the
total number of MPCs where $T_{DS}$ is the channel delay spread. Note that, in
order to achieve high data-rate communications, the channel delay spread is
assumed significantly larger than one symbol duration. Hence, the received
signal encounters severe ISI.

Assuming that the timing is acquired, the received signal can be
expressed as
\begin{equation*}
z(t)= \sum_{k=1}^{K}\sum_{l=0}^{L-1} h_{k,l}x^{(k)}(t-lT_{\tau})+n(t),
\end{equation*} where $n(t)$ is the additive white gaussian noise (AWGN) with
zero mean and a variance of $\sigma_{n}^{2}$. This signal is first passed
through a chip-matched filter (CMF) and then sampled at the chip rate. We
select a total number of $M=(T_{s}+T_{DS})/T_{c}$ observation samples for the
detection of each data bit, where $T_{s}$ is the symbol duration, $T_{DS}$ is
the channel delay spread and $T_{c}$ is the chip duration. Assuming the
sampling starts at the zero-th time instant, then the $m$-th sample can be
expressed as $r_{m}= \int_{mT_{c}}^{(m+1)T_{c}} z(t)p_{r}(t)~ dt,$
where $m=1,2,\dots,M$, $p_{r}(t)=p_{t}^{*}(-t)$ denotes the CMF and
$(\cdot)^{*}$ denotes the complex conjugation. After the chip-rate sampling,
the discrete-time received signal for the $i$-th data bit can be expressed as
$\mathbf r(i)=[r_{1}(i),r_{2}(i),\dots,r_{M}(i)]^{T}$, where $(\cdot)^T$ is the
transposition. We can further express it in a matrix form as
\begin{equation}
\mathbf r(i)=\sum_{k=1}^{K}\sqrt{E_{k}}\mathbf P_{r}\mathbf {H}_{k}\mathbf
P_{t}\mathbf s_{k} b_{k}(i)+\mbox{\boldmath$\eta$}(i)+\mathbf {n}(i),
\label{eq:ri}
\end{equation} where $\mathbf {H}_{k}$ is the Toeplitz channel matrix for the
$k$-th user with the first column being the CIR  of $\mathbf
h_{k}=[h_{k}(0),h_{k}(1),\dots,h_{k}(L-1)]^{T}$ zero-padded to length
$M_{H}=(T_{s}/T_{\tau})+L-1$. Matrix $\mathbf P_{r}$ represents the CMF and
chip-rate sampling with the size $M$-by-$M_{H}$. $\mathbf P_{t}$ denotes the
$(T_{s}/T_{\tau})$-by-$N_{c}$ pulse shaping matrix. The vector
$\mbox{\boldmath$\eta$}(i)$ denotes the ISI from $2G$ adjacent symbols, where
$G$ denotes the minimum integer that is larger than or equal to the scalar term
$T_{DS}/T_{s}$. Here, we express the ISI vector in a general form
that is given by
\begin{equation}
\begin{split}
\mbox{\boldmath$\eta$}(i)&=\sum_{k=1}^{K}\sum_{g=1}^{G}\sqrt{E_{k}}\mathbf
P_{r}\mathbf {H}_{k}^{(-g)}\mathbf P_{t}\mathbf s_{k}
b_{k}(i-g)\\
&+\sum_{k=1}^{K}\sum_{g=1}^{G}\sqrt{E_{k}}\mathbf P_{r}\mathbf
{H}_{k}^{(+g)}\mathbf P_{t}\mathbf s_{k} b_{k}(i+g),
\end{split}
\end{equation} where the channel matrices for the ISI
are given by
\begin{equation}
\mathbf {H}_{k}^{(-g)}=\begin{bmatrix}\mathbf 0 &\mathbf H_{k}^{(u,g)}\\
\mathbf 0& \mathbf 0
\end{bmatrix}~~;~~\mathbf {H}_{k}^{(+g)}=\begin{bmatrix}\mathbf 0 &\mathbf 0\\ \mathbf H_{k}^{(l,g)} & \mathbf 0
\end{bmatrix}.
\end{equation} Note that the matrices $\mathbf H_{k}^{(u,g)}$ and $\mathbf H_{k}^{(l,g)}$ have the same size
as $\mathbf {H}_{k}$, which is $M_{H}$-by-$(T_{s}/T_{\tau})$, and can be
considered as the partitions of an upper triangular matrix $\mathbf H_{\rm up}$
and a lower triangular matrix $\mathbf H_{\rm low}$, respectively, where
\begin{equation*}
\begin{split}
&\mathbf H_{\rm up}=\begin{bmatrix}h_{k}(L-1)     & \dots & h_{k}(L-\frac{T_{DS}-(g-1)T_{s}}{T_{\tau}})\\
                                         & \ddots       &\vdots\\
                                         &        &h_{k}(L-1)
\end{bmatrix}~;\\
&\mathbf H_{\rm low}=\begin{bmatrix}h_{k}(0) & & \\
                                               \vdots  &\ddots   &\\
                                               h_{k}(\frac{T_{DS}-(g-1)T_{s}}{T_{\tau}}-2)&\dots
                                               &h_{k}(0)
\end{bmatrix}.
\end{split}
\end{equation*} These triangular matrices have the row-dimension of
$[T_{DS}-(g-1)T_{s}]/T_{\tau}-1=L-(g-1)T_{s}/T_{\tau}-1$. Note that when the
channel delay spread is large, the row-dimension of these triangular matrices
could surpass the column dimension of the matrix $\mathbf {H}_{k}$, which is
$T_{s}/T_{\tau}$. Hence, in case of
\begin{equation}
L-(g-1)T_{s}/T_{\tau}-1>T_{s}/T_{\tau},~~i.e. ~~~L>gT_{s}/T_{\tau}+1,
\end{equation} the matrix $\mathbf H_{k}^{(u,g)}$ is the last $T_{s}/T_{\tau}$ columns of the
upper triangular matrix $\mathbf H_{\rm up}$ and $\mathbf H_{k}^{(l,g)}$ is the
first $T_{s}/T_{\tau}$ columns of the lower triangular matrix $\mathbf H_{\rm
low}$. When $L<gT_{s}/T_{\tau}+1$, $\mathbf H_{k}^{(u,g)}=\mathbf H_{\rm up}$
and $\mathbf H_{k}^{(l,g)}=\mathbf H_{\rm low}$. It is interesting to review
the expression of the ISI vector via its physical meaning, since the
row-dimension of the matrices $\mathbf H_{k}^{(u,g)}$ and $\mathbf
H_{k}^{(l,g)}$, which is $L-(g-1)T_{s}/T_{\tau}-1$, reflects the time domain
overlap between the data symbol $b(i)$ and the adjacent symbols of $b(i-g)$ and
$b(i+g)$.

In order to estimate the data bit, an $M$-dimensional full-rank filter $\mathbf
w(i)$ can be employed to minimize the MSE cost function:
\begin{equation}
\mathbf J_{\rm MSE}(\mathbf w(i))=E[|d(i)-\mathbf w^{H}(i)\mathbf r(i)|^{2}],
\label{eq:mse}
\end{equation} where  $d(i)$ is the desired signal, $(\cdot)^H$
denotes the Hermitian transpose and $E[\cdot]$ represents the
expected value. Without loss of generality, we consider user 1 as
the desired user and omit the subscript of this user for simplicity.
The optimal solution that minimizes \eqref{eq:mse} is given by
\begin{equation}
\mathbf w_{\rm o}=\mathbf R^{-1} \mathbf p,
\label{eq:mmsesolution-full-rank}
\end{equation} where $\mathbf R=E[\mathbf r(i)\mathbf r^{H}(i)]$ is the correlation matrix of the
discrete-time received signal $\mathbf r(i)$ and $\mathbf p=E[d^{*}(i)\mathbf
r(i)]$ is the cross-correlation vector between $\mathbf r(i)$ and $d(i)$.
The corresponding MMSE can be expressed as:
\begin{equation}
\mathbf {MMSE}_{\rm f}=\sigma^{2}_{d}-\mathbf p^{H}\mathbf
R^{-1}\mathbf p, \label{eq:MMSE-full-rank}
\end{equation}  where $\sigma^{2}_{d}$ is the variance of the desired
signal.
Full-rank adaptive algorithms can update $\mathbf w(i)$ to approach the optimal
solution in \eqref{eq:mmsesolution-full-rank}. The final decision is made by
$\hat {b}(i)={\rm sign}(\rm {\Re}[\mathbf {w}^{H}(i) \mathbf
{r}(i)]),$ where ${\rm sign}(\cdot)$ is the algebraic sign function and
$\rm{\Re} (\cdot)$ represents the real part of a complex number.
The full-rank adaptive filters experience slow
convergence rate in DS-UWB systems because of the long channel delay
spread. In order to accelerate the convergence and increase the
robustness against interference, we propose a generic reduced-rank
scheme in what follows.



\section{Generic Reduced-Rank Scheme and Problem Statement}
\label{sec:rrscheme}

Reduced-rank signal processing can be divided into two parts: an
$M$-by-$D$ projection matrix that projects the $M$-dimensional
received signal onto a $D$-dimensional subspace (where $D\ll M$),
and a $D$-dimensional reduced-rank linear filter that produces the
output. The projection stage of the reduced-rank schemes is given by
\begin{equation}
\bar{\mathbf r}(i)=\mathbf T ^ {H}(i) \mathbf r(i),
\end{equation} where $\bar{\mathbf r}(i)$ is the reduced-rank signal and $\mathbf T(i)$ is the projection
matrix that can be expressed as
\begin{equation}
\mathbf
{T}(i)=[\mbox{\boldmath$\phi$}_{1}(i),\dotsb,\mbox{\boldmath$\phi$}_{d}(i),\dotsb
\mbox{\boldmath$\phi$}_{D}(i)],
\end{equation} where $\{ \mbox{\boldmath$\phi$}_{d}(i)|$ $d=1, \dots, D\}$ are the $M$-dimensional basis
vectors. The vector $\bar{\mathbf r}(i)$ is then passed through a
$D$-dimensional linear filter. The MMSE solution of such a filter is
\begin{equation}
\bar{\mathbf w}_{\rm o}=\bar{\mathbf R} ^ {-1} \bar{\mathbf p},
\label{eq:woptfullrank}
\end{equation} where $\bar{\mathbf R}=E[\bar{\mathbf r}(i)\bar{\mathbf
r}^{H}(i)]$ and $\bar{\mathbf p}=E[d^{*}(i)\bar{\mathbf r}(i)]$.

In reduced-rank schemes, the main challenge is how to effectively design the
projection matrix $\mathbf T(i)$.
In order to simplify the expression of the proposed SAABF scheme in
later sections, the reduced-rank signal is expressed as
\begin{equation}
\begin{split}
&\bar{\mathbf r}(i)=\mathbf T ^ {H}(i) \mathbf r(i)\\
&=\begin{bmatrix}
\mathbf r^{T}(i)        & & &\\
                        & \mathbf r^{T}(i) & & \\
                        &                  &\ddots  &\\
                        &                  &        &\mathbf r^{T}(i)\\
\end{bmatrix}_{D\times MD}
\begin{bmatrix}
\mbox{\boldmath$\phi$}_{1}(i)\\
\mbox{\boldmath$\phi$}_{2}(i)\\
\vdots\\
\mbox{\boldmath$\phi$}_{D}(i)\\
\end{bmatrix}^{*}_{MD\times 1}\\
&=\mathbf R_{\rm in}(i)\mathbf t(i), \label{eq:generalrrsignal}
\end{split}
\end{equation} where the projection matrix is transformed into a vector form, and $\mathbf t(i)$ is called
projection vector in what follows. It can be shown that the $d$-th
element in the reduced-rank signal is $\bar{r}_{d}(i)=\mathbf
r^{T}(i)\mbox{\boldmath$\phi$}_{d}^{*}(i)$, where $d=1, \dots, D$.
The generic reduced-rank scheme is proposed to jointly
and iteratively adapt the projection vector and the reduced-rank
linear estimator to minimize the MSE cost function
\begin{equation}
\begin{split}
\mathbf J_{\rm MSE}(\bar{\mathbf w}(i),\mathbf
t(i))&=E[|d(i)-\bar{\mathbf w}^{H}(i)\mathbf R_{\rm in}(i)\mathbf
t(i)|^{2}]. \label{eq:costfuntionfullranknew}
\end{split}
\end{equation}
The MMSE solution of the reduced-rank filter in the generic scheme has the same
form as \eqref{eq:woptfullrank}. By setting the gradient vector of
\eqref{eq:costfuntionfullranknew} with respect to
$\mathbf t(i)$ to a null vector, the optimum projection vector is given by
\begin{equation}
\begin{split}
&\mathbf t_{\rm opt}={\mathbf R}_{w}^{-1} {\mathbf p}_{w}.\\
\end{split}
\label{eq:mmsesolution-generic}
\end{equation} where ${\mathbf R}_{w}=E[\mathbf R_{\rm in}^{H}(i)\bar{\mathbf
w}(i)\bar{\mathbf w}^{H}(i)\mathbf R_{\rm in}(i)]$ and ${\mathbf
p}_{w}=E[d(i)\mathbf R_{\rm in}^{H}(i)\bar{\mathbf w}(i)]$. The MMSE of the
generic scheme can be expressed as:
\begin{equation}
\mathbf {MMSE}_{\rm g}=\sigma^{2}_{d}-\bar{\mathbf
p}^{H}\bar{\mathbf R}^{-1}\bar{\mathbf p}.
\end{equation}
Note that when adaptive algorithms are implemented to estimate
$\bar{\mathbf w}_{\rm o}$ and $\mathbf t_{\rm opt}$, $\bar{\mathbf
w}(i)$ is a function of $\mathbf t(i)$ and $\mathbf t(i)$ is a
function of $\bar{\mathbf w}(i)$. Thus, the joint MMSE design is not
in a closed form and one possible solution for such optimization
problem is to jointly and iteratively adapt these two parts with an
initial guess. The joint-adaptation is operated as follow: for the
$i$-th time instant, $\bar{\mathbf w}(i)$ is obtained with the
knowledge of $\mathbf t(i-1)$ and $\bar{\mathbf w}(i-1)$, then
$\mathbf t(i)$ is updated with $\mathbf t(i-1)$ and $\bar{\mathbf
w}(i)$. The iterative-adaptation is to repeat the joint-adaptation
until the satisfactory estimates are obtained. Hence, the number of
iterations are environment dependent. Compared with existing
reduced-rank schemes such as MSWF \cite{MLHonig2001} and AVF
\cite{dapados2001}, this generic scheme enables the projection
vector and the reduced-rank filter to exchange information at each
iteration. This feature leads to a more effective operation of the
adaptive algorithms. However, the drawback of such a
feature is that we cannot obtain a closed form design.
It will be illustrated by the simulation results that
this generic scheme outperforms the MSWF \cite{MLHonig2001} and AVF
\cite{dapados2001} with a few iterations.

Note that in DS-UWB systems where the length of the full-rank received signal
$M$ is large, the complexity of updating the $MD$-dimensional projection vector
is very high. In order to reduce the complexity of this generic scheme,we
propose the following SAABF scheme. 

\section{Proposed SAABF Scheme and Filter Design}
\label{sec:proposedscheme}
\begin{figure*}[htb]
\begin{minipage}[h]{1.0\linewidth}
  \centering
  \centerline{\epsfig{figure=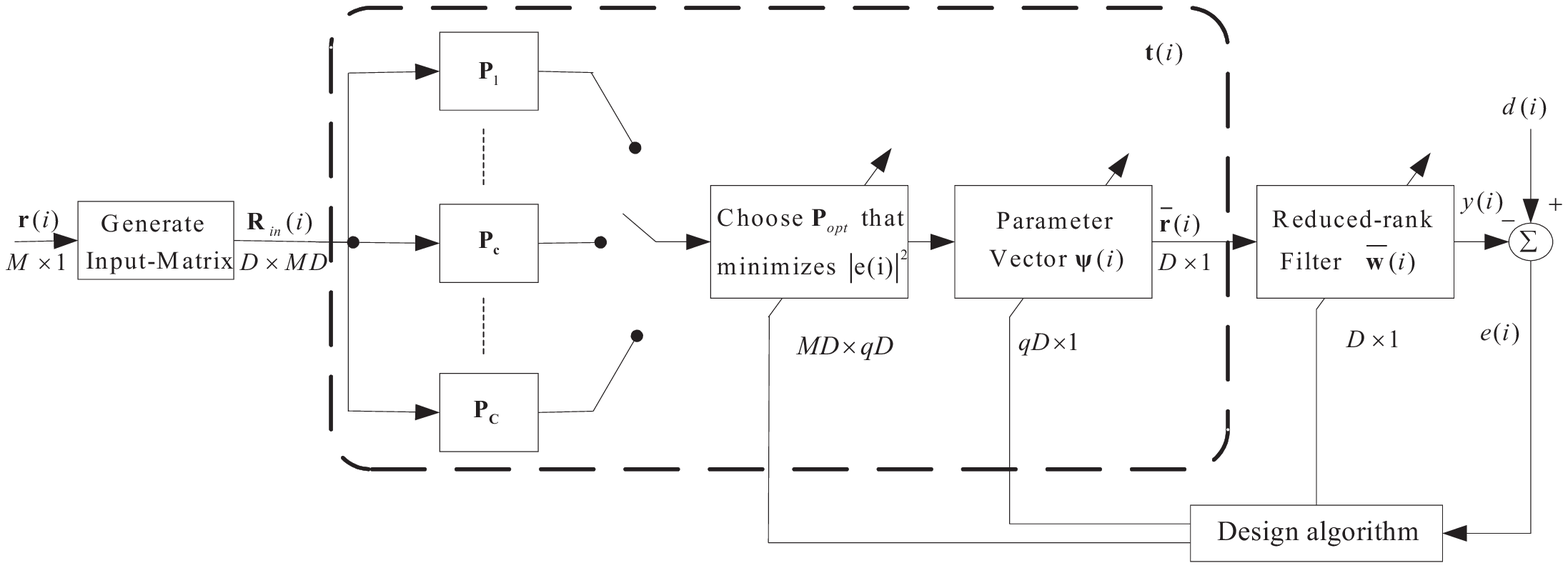,scale=0.8}}
\end{minipage}
\caption{Block diagram of the proposed reduced-rank linear receiver
using the SAABF scheme.} \label{fig:schememodel}
\end{figure*}
In this section we detail the proposed switched approximation of adaptive basis
functions (SAABF) scheme, whose primary idea is to constrain the structure of
the $MD$-dimensional projection vector $\mathbf t(i)$, using a multiple-branch
framework such that the number of coefficients to be computed is substantially
reduced. The block diagram of the proposed SAABF scheme is shown in
Fig.\ref{fig:schememodel}. There are $C$ branches in the SAABF scheme. For each
branch, a projection vector is equivalent to a projection matrix
$\mathbf T_c(i)=[\mbox{\boldmath$\phi$}_{c,1}(i),\dotsb
,\mbox{\boldmath$\phi$}_{c,d}(i),\dotsb \mbox{\boldmath$\phi$}_{c,D}(i)],$
where $c=[1, 2, \dotsb, C]$, $d=[1, 2, \dotsb,D]$ and the $M$-dimensional
adaptive basis function is given by
\begin{equation}
\mbox{\boldmath$\phi$}_{c,d}(i)=
\begin{bmatrix}
\mathbf {0}_{z_{c,d}\times q}\\
\mathbf {I}_q\\
\mathbf {0}_{(M-q-z_{c,d})\times q}\\
\end{bmatrix}_{M \times q}{\mbox{\boldmath$\varphi$}_{d}}(i)=\mathbf {Z}_{c,d} {\mbox{\boldmath$\varphi$}_{d}}(i),
\label{eq:forzcd}
\end{equation} where $z_{c,d}$ is the number of zeros before the $q$-by-$1$ function ${\mbox{\boldmath$\varphi$}_{d}}(i)$
 (where $q\ll M$),
which is called the inner function in what follows. The matrix
$\mathbf {Z}_{c,d}$ consists of zeros and ones. With an $q$-by-$q$
identity matrix $\mathbf I_{q}$ in the middle, the zero matrices
have the size of $z_{c,d}$-by-$q$ and $(M-q-z_{c,d})$-by-$q$,
respectively.
Hence, we can express the projection vector as
\begin{equation}
\begin{split}
\mathbf t_{c}(i)&=\begin{bmatrix} \mbox{\boldmath$\phi$}^{T}_{c,1}(i),
\mbox{\boldmath$\phi$}^{T}_{c,2}(i), \cdots,
\mbox{\boldmath$\phi$}^{T}_{c,D}(i)
\end{bmatrix}^{H}\\
&=
\begin{bmatrix}
\mathbf {Z}_{c,1}    &               &       &\\
               &\mathbf {Z}_{c,2}    &       &\\
               &               &\ddots       &\\
               &               &             &\mathbf {Z}_{c,D} \\
\end{bmatrix}\begin{bmatrix}
{\mbox{\boldmath$\varphi$}_{1}}(i)\\
{\mbox{\boldmath$\varphi$}_{2}}(i)\\
\vdots\\
{\mbox{\boldmath$\varphi$}_{D}}(i)\\
\end{bmatrix}^{*}=\mathbf P_c {\mbox{\boldmath$\psi$}} (i),
\end{split}
\label{eq:t_c}
\end{equation} where the $MD$-by-$qD$ block diagonal matrix $\mathbf P_c$ is called position matrix which
determines the positions of the $q$-dimensional inner functions and
${\mbox{\boldmath$\psi$}} (i)$ denotes the $qD$-dimensional
projection vector which is constructed by the inner functions. For
each time instant, the rank-reduction in the SAABF scheme is
achieved by selecting the position matrix $\mathbf P(i)$
instantaneously from a set of pre-stored position matrices $\mathbf
P_{c}$, where $c=1, \dots, C$, and updating the
${\mbox{\boldmath$\psi$}}(i)$. Compared with
\eqref{eq:generalrrsignal}, equation \eqref{eq:t_c} shows the
constraint we use in the SAABF scheme. With the multi-branch
structure, the dimension of the projection vector is shortened from
$MD$ to $qD$. 

For simplicity, we denote the proposed scheme with its main parameters as
'SAABF (C,D,q)', where $C$ is the number of branches, $D$ is the length of the
reduced-rank filter and $q$ is the length of the inner function. Note that in
the case of the SAABF (1,D,M), where $C=1$ and $q=M$, the proposed scheme is
equivalent to the generic scheme described in Section \ref{sec:rrscheme}. For
the SAABF (1,1,M), where $C=1$, $D=1$ and $q=M$, the proposed scheme can be
considered to be a full-rank scheme. All these equivalences are proved in the
appendix \ref{app:equivalence}, which shows the optimal solutions in these
scenarios will lead to the same MMSE.

It is interesting to note that the adaptation in the proposed SAABF scheme can
be considered a hybrid adaptive technique, which includes a discrete parameter
optimization for choosing the instantaneous position matrix and a continuous
filter design for adapting the projection vector and the reduced-rank filter.
In what follows, we detail the discrete parameter optimization and
the filter design. 
\subsection{Discrete Parameter Optimization}
\label{subsec:dpo}
In this section, the selection rule for choosing $\mathbf P(i)$ is introduced
and the designs of the pre-stored position matrices $\mathbf P_{c}$ are
detailed. The problem of computing the optimal $\mathbf P(i)$ is a discrete
optimization problem since $\mathbf P(i)$ can be considered as a time
independent parameter which is selected from a set of pre-stored matrices at
each time instant for minimizing the instantaneous squared error. The output
signal of each branch is given by
\begin{equation*}
y_{c}(i)=\bar {\mathbf w}^{H}(i)\mathbf R_{\rm in}(i)\mathbf t_{c}(i)=\bar
{\mathbf{w}}^{H}(i) \mathbf R_{\rm in}(i)\mathbf P_{c}
{\mbox{\boldmath$\psi$}}(i),
\end{equation*} where the corresponding error signal is $e_{c}(i)=d(i)-y_{c}(i)$. Hence, the
selection rule can be expressed as
\begin{equation}
c_{\rm opt}=\arg\min_{c\in\{1,\ldots,C\}}
|e_{c}(i)|^{2},~e(i)=e_{c_{\rm opt}}(i),~\mathbf P(i)=\mathbf P_{c_{\rm opt}}.\\
\label{eq:copt}
\end{equation}
As shown in \eqref{eq:forzcd} and \eqref{eq:t_c}, the position matrices are
distinguished by the values of $z_{c,d}$. The optimal way for selecting
$z_{c,d}$ is to test all the possibilities of the position matrices and choose
a structure which corresponds to the minimum squared error. However, in the
DS-UWB system, the number of possible positions is $(M-q)^{D}$, where $M$ is
much larger than $q$ and $D$, say $M=112$ and $q=D=3$. Therefore, it is too
expensive to find the optimal position matrix from such a huge number of
possibilities. Hence, we design a small number of $C$ pre-stored position
matrices that enables us to find a sub-optimum instantaneous position matrix
that provides an attractive tradeoff between performance and complexity. 
Note that the number $C$ can be considered as a system parameter
for the designer, increasing the number of position matrices will
benefit the performance but also increase the complexity. 
In section \ref{sec:branchnumber}, we propose a branch number selection
algorithm to determine the C within a given range to decrease the averaged
required number of branches while maintaining the performance.

For designing the pre-stored matrices, we propose a simple deterministic way to
set the values of $z_{c,d}$ as follows
\begin{equation}
z_{c,d}=\lfloor \frac{M}{D} \rfloor \times (d-1)+(c-1)q,
\end{equation}where $c=1, \dots, C$ and $d=1, \dots, D$. Bearing in mind the matrix form shown in  \eqref{eq:forzcd} and
\eqref{eq:t_c}, the first $MD$-by-$qD$ position matrix $\mathbf P_{1}$ can be
expressed as
\begin{equation}
\mathbf P_{1}=
\begin{bmatrix}
\mathbf {I}_{q}             &                                              & & \\
\mathbf {0}_{M-q} &                                                        &  &\\
                            & \mathbf {0}_{\lfloor \frac{M}{D} \rfloor}    &  &\\
                            & \mathbf {I}_{q}                              &  &\\
                            & \mathbf {0}_{M-q-\lfloor \frac{M}{D} \rfloor}&  &\\
                            &&\ddots&\\
                            &&&\mathbf {0}_{\lfloor \frac{M}{D} \rfloor (D-1)}\\
                            &&&\mathbf {I}_{q}\\
                            &&&\mathbf {0}_{M-q-\lfloor \frac{M}{D} \rfloor (D-1)}\\
\end{bmatrix},
\label{eq:positionmatrixp1}
\end{equation} where all the zero and identity matrices have $q$ columns and the subscripts denote the number of
rows of these matrices. We remark that the proposed approach arranges the
$q$-by-$q$ identity matrices in a simple fixed sliding pattern. This then
allows efficient generation of the remaining position matrices. For example,
the second projection matrix $\mathbf P_{2}$ can be considered as a shifted
version of $\mathbf P_{1}$, in which each column has been shifted down by $q$
elements. It should be noted that the pre-stored position matrices
can also be generated randomly, in which approach the values of $z_{c,d}$ are
set randomly. However, the random method will require extra storage space for
all the pre-stored matrices and the performance of this method is inferior to
the proposed deterministic method.

\subsection{Filter Design}

After determining the position matrix $\mathbf P(i)$, 
%
%
the LS design of the reduced-rank filter and the projection vector
can be developed to minimize the following cost function
\begin{equation}
\mathbf J_{\rm LS}(\bar
{\mathbf{w}}(i),{\mbox{\boldmath$\psi$}}(i))=\sum_{j=1}^{i}\lambda^{i-j}|d(j)-\bar
{\mathbf{w}}^{H}(i) \mathbf R_{\rm in}(j)\mathbf P(i)
{\mbox{\boldmath$\psi$}}(i)|^{2}, \label{eq:costfunctionLS}
\end{equation}where $\lambda$ is a forgetting factor. 
Firstly, we calculate the gradient of \eqref{eq:costfunctionLS} with
respect to $\bar{\mathbf w}(i)$, which is
\begin{equation}
{\mathbf g_{\rm LS}}_{\bar{w}^{*}(i)}=-\bar{\mathbf{p}}_{w_{\rm
LS}}(i)+ \bar{\mathbf{R}}_{w_{\rm LS}}(i)\bar
{\mathbf{w}}(i),\label{eq:gLS}
\end{equation} where $\bar{\mathbf{p}}_{w_{\rm LS}}(i)=\sum_{j=1}^{i}\lambda^{i-j}d^{*}(j)\bar {\mathbf
r}(j)$ and $\bar{\mathbf{R}}_{w_{\rm
LS}}(i)=\sum_{j=1}^{i}\lambda^{i-j}\bar{\mathbf r}(j)\bar{\mathbf
r}^{H}(j)$. Assuming that ${\mbox{\boldmath$\psi$}}(i)$ is fixed,
the LS solution of the reduced-rank filter is
\begin{equation}
\bar{\mathbf w}_{\rm LS}(i)=\bar{\mathbf{R}}_{w_{\rm
LS}}^{-1}(i)\bar{\mathbf{p}}_{w_{\rm LS}}(i). \label{eq:LSsolutionw}
\end{equation}
Secondly, we examine the gradient of \eqref{eq:costfunctionLS} with
respect to ${\mbox{\boldmath$\psi$}}(i)$, which is
\begin{equation}
{\mathbf g_{\rm
LS}}_{\psi^{*}(i)}=-{\mathbf{p}}_{\psi_{LS}}(i)+{\mathbf{R}}_{\psi_{\rm
LS}}(i){\mbox{\boldmath$\psi$}}(i), \label{eq:gphiLS}
\end{equation} where ${\mathbf{p}}_{\psi_{\rm LS}}(i)=\sum_{j=1}^{i}\lambda^{i-j}d(j)\mathbf
r_{\psi}(j)$, ${\mathbf{R}}_{\psi_{\rm
LS}}(i)=\sum_{j=1}^{i}\lambda^{i-j}\mathbf r_{\psi}(j)\mathbf
r_{\psi}^{H}(j){\mbox{\boldmath$\psi$}}(i)$ and $\mathbf r_{\psi}(j)=\mathbf
P^{H}(j)\mathbf R_{\rm in}^{H}(j)\bar{\mathbf w}(j)$. With the assumption that
$\bar {\mathbf{w}}(i)$ is fixed, the LS solution of the projection vector is
\begin{equation}
{\mbox{\boldmath$\psi$}}_{\rm LS}(i)={\mathbf{R}}_{\psi_{\rm
LS}}^{-1}(i){\mathbf{p}}_{\psi_{\rm LS}}(i).\label{eq:LSsolutionphi}
\end{equation}
Finally, \eqref{eq:LSsolutionw} and \eqref{eq:LSsolutionphi}
summarize the LS design of the reduced-rank filter and the
projection vector in the SAABF scheme. A discussion on the
optimization of the SAABF scheme is presented in appendix
\ref{app:optdiscussion}.

\section{Adaptive Algorithms}
\label{sec:adaptivealgorithms}

In this section, joint LMS and RLS algorithms are developed for
estimating the reduced-rank filter and the projection vector. The
complexity analysis is also given to compare the computational load
of existing and the proposed algorithms. We remark that in the SAABF
scheme, when a number of branches are implemented, the joint
adaptation only requires one iteration for each time instant.

\subsection{LMS Version}

The joint LMS version of the SAABF scheme is developed to minimize
the MSE cost function:
\begin{equation}
\mathbf {J}_{\rm MSE}(\bar
{\mathbf{w}}(i),{\mbox{\boldmath$\psi$}}(i)) =E[|d(i)-\bar
{\mathbf{w}}^{H}(i)\mathbf R_{\rm in}(i)\mathbf P(i)
{\mbox{\boldmath$\psi$}}(i)|^{2}],\label{eq:mmse-cf}
\end{equation} where $\mathbf P(i)$ is the instantaneous position matrix.
The MMSE solution of the SAABF scheme is shown in the appendix
\ref{app:equivalence}.

At the $i$-th time instant, we firstly determine the instantaneous position
matrix with the selection rule \eqref{eq:copt}. Then, the reduced-rank filter weight vector $\bar {\mathbf{w}}(i)$ can be updated
with the LMS algorithm \cite{haykin}. 
Taking the gradient vector of \eqref{eq:mmse-cf} with
respect to $\bar {\mathbf{w}}(i)$ and using the instantaneous values
of the gradient vector, the adaptation equation for the reduced-rank
filter is given by
\begin{equation}
\bar {\mathbf{w}}(i+1)=\bar {\mathbf{w}}(i)+\mu_{w} \mathbf R_{\rm
in}(i)\mathbf P(i) {\mbox{\boldmath$\psi$}}(i) e^{*}(i),
\label{eq:lmsadaptw}
\end{equation} where $\mu_{w}$ is the step size.
With the knowledge of the updated reduced-rank filter, the
projection vector can be adapted to minimize the cost function
\eqref{eq:mmse-cf}. 
Taking the gradient vector of \eqref{eq:mmse-cf} with respect to
${\mbox{\boldmath$\psi$}}(i)$ and using the instantaneous estimate of the
gradient vector, the adaptation equation for the projection vector is obtained
as
\begin{equation}
{\mbox{\boldmath$\psi$}}(i+1)={\mbox{\boldmath$\psi$}}(i)+\mu_{\psi}\mathbf
P^{H}(i)\mathbf R_{\rm in}^{H}(i)\bar {\mathbf{w}}(i+1)e(i),
\label{eq:lmsadaptphi}
\end{equation} where $\mu_{\psi}$ is the step size. We summarize the
LMS version of the SAABF scheme in Table \ref{tab:tablelms}.

\subsection{RLS Version}
Let us consider the RLS design of the SAABF scheme, which can be
developed to minimize the cost function shown in
\eqref{eq:costfunctionLS}. The instantaneous position matrix is
determined with the selection rule \eqref{eq:copt}.
The reduced-rank filter will be updated first. The gradient of
\eqref{eq:costfunctionLS} with respect to $\bar{\mathbf w}(i)$ is shown in
\eqref{eq:gLS}. By applying the matrix inversion lemma to $\bar{\mathbf{R}}_{w_{\rm LS}}(i)$,
we obtain its inverse matrix in a recursive way as
\begin{equation}
\bar{\mathbf{R}}_{w_{\rm
LS}}^{-1}(i)=\lambda^{-1}\bar{\mathbf{R}}_{w_{\rm
LS}}^{-1}(i-1)-\lambda^{-1}\mathbf K_{w}(i)\bar{\mathbf
r}^{H}(i)\bar{\mathbf{R}}_{w_{\rm
LS}}^{-1}(i-1),\label{eq:inverseRforRLS-w}
\end{equation}where $\mathbf K_{w}(i)= \big(\bar{\mathbf{R}}_{w_{\rm
LS}}^{-1}(i-1)\bar{\mathbf r}(i)\big)/\big(\lambda+\bar{\mathbf
r}^{H}(i)\bar{\mathbf{R}}_{w_{\rm LS}}^{-1}(i-1)\bar{\mathbf r}(i)\big).$
In order to obtain a recursive update equation, we express the
vector $\bar {\mathbf{p}}_{w_{\rm LS}}(i)$ as
\begin{equation}
\bar {\mathbf{p}}_{w_{\rm LS}}(i)=\lambda \bar {\mathbf{p}}_{w_{\rm
LS}}(i-1)+d^{*}(i)\bar {\mathbf r}(i).\label{eq:Pforrls-w}
\end{equation}
By substituting  \eqref{eq:inverseRforRLS-w} and \eqref{eq:Pforrls-w} into
\eqref{eq:gLS} and setting the gradient to zero, we obtain the RLS adaptation
equation for the reduced-rank filter as
\begin{equation}
\bar {\mathbf{w}}(i)= \bar {\mathbf{w}}(i-1)+\mathbf K_{w}(i)
e^{*}(i).
\end{equation}
With the knowledge of the updated reduced-rank filter, we can adapt
the projection vector to minimize the cost function
\eqref{eq:costfunctionLS}. The gradient of \eqref{eq:costfunctionLS}
with respect to ${\mbox{\boldmath$\psi$}}(i)$ is shown in
\eqref{eq:gphiLS}.

In order to obtain the recursive update equation for the projection vector, we
express ${\mathbf{p}}_{\psi_{\rm LS}}(i)$ in a recursive form as:
\begin{equation}
{\mathbf{p}}_{\psi_{\rm LS}}(i)=\lambda {\mathbf{p}}_{\psi_{\rm
LS}}(i-1)+d(i)\mathbf r_{\psi}(i),\label{eq:Pforrls-phi}
\end{equation} where $\mathbf r_{\psi}(j)=\mathbf P^{H}(j)\mathbf R_{\rm
in}^{H}(j)\bar{\mathbf w}(j)$.

Applying the matrix inversion lemma to ${\mathbf{R}}_{\psi_{\rm
LS}}(i)$, we obtain its inverse recursively
\begin{equation}
{\mathbf{R}}_{\psi_{\rm
LS}}^{-1}(i)=\lambda^{-1}{\mathbf{R}}_{\psi_{\rm
LS}}^{-1}(i-1)-\lambda^{-1}\mathbf K_{\psi}(i)\mathbf
r_{\psi}^{H}(i){\mathbf{R}}_{\psi_{\rm LS}}^{-1}(i-1),
\label{eq:inverseRforRLS-phi}
\end{equation} where $\mathbf K_{\psi}(i)= \big({\mathbf{R}}_{\psi_{\rm
LS}}^{-1}(i-1)\mathbf r_{\psi}(i)\big)/\big(\lambda+\mathbf
r_{\psi}^{H}(i){\mathbf{R}}_{\psi_{\rm LS}}^{-1}(i-1)\mathbf
r_{\psi}(i)\big).$
By substituting  \eqref{eq:Pforrls-phi} and
\eqref{eq:inverseRforRLS-phi} into \eqref{eq:gphiLS} and setting the
gradient to zero, we obtain the RLS adaptation equation for the
projection vector
\begin{equation}
{\mbox{\boldmath$\psi$}}(i)= {\mbox{\boldmath$\psi$}}(i-1)+\mathbf
K_{\psi}(i)e(i).
\end{equation}
The RLS version of the SAABF scheme is summarized in Table
\ref{tab:tablelms}.
\begin{table}[h]
\centering
 \caption{\normalsize proposed adaptive algorithms.} {
\begin{tabular}{ll}
\hline\hline
LMS : \rule{0pt}{2.6ex} \rule[-1.2ex]{0pt}{0pt}&\\
\hline Step 1:&Initialization:\rule{0pt}{2.6ex} \\
& ${\mbox{\boldmath$\psi$}}(0)$=$\rm{ones}(qD,1)$ and $\bar
{\mathbf{w}}(0)$=$\rm{zeros}(D,1)$\\
& Set values for $\mu_{w}$ and $\mu_{\psi}$\\
& Generate the position matrices $\mathbf P_{1}$, \ldots, $\mathbf P_{C}$\\

Step 2:&For i=0, 1, 2, \ldots.\\
&
\begin{tabular}{ll}
 (1) Compute the error signals $e_{c}(i)$ for each branch,\\
 (2) Select the branch $c_{\rm opt}=\arg \min_{c\in\{1,\ldots,C\}} |e_{c}(i)|^{2}$, \\
 (3) Set the instantaneous position matrix   $\mathbf P(i)$=$\mathbf P_{c_{\rm opt}}$,\\
 (4) Update $\bar {\mathbf{w}}(i+1)$ using \eqref{eq:lmsadaptw}\\
 (5) Update ${\mbox{\boldmath$\psi$}}(i+1)$  using \eqref{eq:lmsadaptphi}.\\
\end{tabular}
\end{tabular}
\begin{tabular}{ll}
RLS :&\rule{0pt}{2.6ex} \rule[-1.2ex]{0pt}{0pt}\\
\hline Step 1:&Initialization:\rule{0pt}{2.6ex}\\
& ${\mbox{\boldmath$\psi$}}(0)$=$\rm{ones}(qD,1)$ and $\bar
{\mathbf{w}}(0)$=$\rm{zeros}(D,1)$\\
& $\bar{\mathbf{R}}_{w_{\rm LS}}^{-1}(0)$=$\mathbf I_{D}/\delta_{w}$ and ${\mathbf{R}}_{\psi_{\rm LS}}^{-1}(0)$=$\mathbf I_{qD}/\delta_{\psi}$\\
& Set values for $\lambda$, $\delta_{w}$ and $\delta_{\psi}$\\
& Generate the position matrices $\mathbf P_{1}$, \ldots, $\mathbf P_{C}$\\

Step 2:&For i=1, 2, \ldots.\\
&
\begin{tabular}{ll}
 (1) Compute the error signals $e_{c}(i)$ for each branch,\\
 (2) Select the branch $c_{\rm opt}=\arg \min_{c\in\{1,\ldots,C\}} |e_{c}(i)|^{2}$, \\
 (3) Set the instantaneous position matrix   $\mathbf P(i)$=$\mathbf P_{c_{\rm opt}}$,\\
 (4) Update $\bar {\mathbf{w}}(i)= \bar {\mathbf{w}}(i-1)+\mathbf K_{w}(i) e^{*}(i)$,\\
 (5) Update $\bar{\mathbf{R}}_{w_{\rm LS}}^{-1}(i)$ using \eqref{eq:inverseRforRLS-w},\\
 (6) Update ${\mbox{\boldmath$\psi$}}(i)= {\mbox{\boldmath$\psi$}}(i-1)+\mathbf
K_{\psi}(i)e(i)$,\\
 (7) Update ${\mathbf{R}}_{\psi_{\rm LS}}^{-1}(i)$ using \eqref{eq:inverseRforRLS-phi}.\\
   \hline
\end{tabular}
\end{tabular}
 } \label{tab:tablelms}
\end{table}
\subsection{Complexity Analysis}
\begin{table*} \centering \caption{\normalsize Complexity analysis}
\begin{tabular}{l l l}
\hline\hline Algorithm & Complex Additions & Complex Multiplications \rule{0pt}{2.6ex} \rule[-1.2ex]{0pt}{0pt}\\
 \hline
Full-Rank LMS & $2M$ & $2M+1$ \rule{0pt}{2.6ex}\\
Full-Rank RLS & $3M^{2}+M$ & $4(M^{2}+M)$ \\
MSWF-LMS & $DM^{2}+(D+2)M$ & $(D+1)M^{2}+(3D+2)M+2D+1$ \\
MSWF-RLS & $DM^{2}+(D+2)M+3D^{2}-D$ & $(D+1)M^{2}+(3D+2)M+4(D^{2}+D)$ \\
AVF      & $(3D+1)M^{2}+M-2D-1$             & $(5D+2)M^{2}+(D+1)M$\\
SAABF(C,D,q)-LMS & $qD(C+1)-CD+C+D$ & $DM+2Dq(C+1)+D+2$\\
SAABF(C,D,q)-RLS & $4(qD)^{2}+CD(q-1)+3D^{2}+C+D$ & $DM+5(qD)^{2}+2CDq+4D^{2}+3Dq+3D$\\
\hline
\end{tabular}
\label{tab:Complexity analysis}
\end{table*}
\begin{figure}[htb]
\begin{minipage}[b]{1.0\linewidth}
  \centering
  \centerline{\epsfig{figure=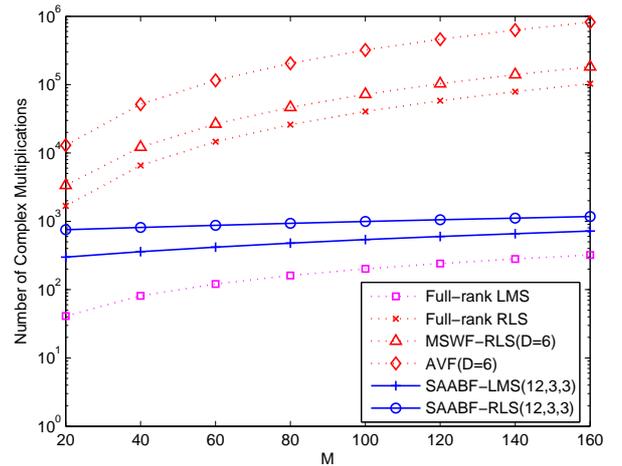,scale=0.6}}
\end{minipage}
\caption{The computational complexity of the linear adaptive algorithms.}
\label{fig:complexity}
\end{figure}
The computational complexity for different adaptive algorithms with
respect to the number of complex additions and complex
multiplications for each processed data bit is shown in Table
\ref{tab:Complexity analysis}. We compare the complexity of the
full-rank LMS and RLS, the LMS and RLS versions of the MSWF, the AVF
and the proposed SAABF scheme. The quantity $M$ is the length of the
full-rank filter, D is the dimension of the subspace, $C$ is the
number of branches in the SAABF scheme and $q$ is the length of the
inner function. 
In Fig.\ref{fig:complexity}, the number of complex multiplications
of the linear adaptive algorithms are shown as a function of $M$. We
remark that the complexity of the receiver with the proposed SAABF
scheme is linearly proportional to the length of the received signal
and is much lower than the existing reduced-rank schemes in the
large signal length scenarios. It should be noted that for each time
instant the SAABF scheme requires one simple search procedure, which
will select the minimum squared-error from a C-dimensional error
vector.

There is an extremely simple configuration of the proposed scheme
that can be expressed as SAABF (C,D,1), in which the length of the
inner function is only $1$ and the projection vector
${\mbox{\boldmath$\psi$}}(i)$ is fixed to its initial value of
${\mbox{\boldmath$\psi$}}(i)=\rm{ones}(D,1)$. This feature
significantly reduces the complexity of the SAABF scheme and the
performance of this configuration will be illustrated with
simulation results.


\section{Model Order and Parameter Adaptation}
\label{sec:parameters}

In the SAABF (C,D,q) scheme, the computational complexity and the
performance are highly dependent on the values of the parameter $C$
and the model order $D$ and $q$. Although we can set suitable values
for these parameters in a specific operation environment with some
performance requirements, the best tradeoffs between the complexity
and performance usually can not be obtained. In order to choose
these parameters automatically and effectively in different
environments, we propose adaptive algorithms as follows.

\subsection{Branch Number Selection}
\label{sec:branchnumber}

The algorithm for selecting the sub-optimum branch number is
developed with the observations: all the branches will be used at
least once but there are some branches that are more likely to be
selected; for a target squared-error, with a given number of
branches, it is unnecessary to test all of them at each time
instant, we can choose the first one that assures the target. With
these observations and assuming that $D$ and $q$ are fixed, we
propose an algorithm to select the number of branches. Firstly, we
set a minimum and a maximum number of branches, denoted as $C_{\rm
min}$ and $C_{\rm max}$, respectively. Then, we define a threshold
$\gamma$ that is related to the MMSE. For each time instant, we test
the first $C_{\rm min}$ branches, if the MSE target is not assured,
we test the $(C_{\rm min}+1)$-th branch and so on. We stop the
search when the target is achieved or the maximum allowed number of
branches $C_{\rm max}$ is reached. The proposed algorithm can be
expressed as
\begin{equation}
C_{\rm r}(i)=\arg \min_{c\in\{C_{\rm min},\ldots,C_{\rm max}\}}
[|e_{c}^{2}(i)-e^{2}_{\rm MMSE}|<\gamma], \label{eq:cselectioneq}
\end{equation} where $e_{c}(i)=d(i)-\bar {\mathbf{w}}^{H}(i) \mathbf R_{\rm in}(i)\mathbf
P_{c} {\mbox{\boldmath$\psi$}}(i)$ is the error signal corresponding to the
$c$-th branch and $C_{\rm r}(i)$ represents the required number of branches at
the $i$-th time instant. Note that the $e_{\rm MMSE}$ is the ideal minimum
error signal and we can replace it with a given value for the target
environment. The aim of this selection algorithm is to reduce the average
number of used branches while maintaining the BER (or MSE) performance.

\subsection{Rank Adaptation}
The computational complexity and the performance of the novel SAABF
reduced-rank scheme is sensitive to the determined rank $D$. Unlike prior work
that used the approach proposed in \cite{MLHonig2002}, we develop a rank
adaptation algorithm based on the \textit{a posteriori} LS cost function to
estimate the MSE, which is a function of the parameters $\bar
{\mathbf{w}}_{D}^{H}(i)$, $\mathbf R_{{\rm in},D}(i)$, $\mathbf P_{D}(i)$ and $
{\mbox{\boldmath$\psi$}}_{D}(i)$
\begin{equation}
\mathscr{C}_{D}(i)=\sum^{i}_{n=0} \lambda_{D}^{i-n}|d(i)-\bar
{\mathbf{w}}_{D}^{H}(i) \mathbf R_{{\rm in},D}(i)\mathbf P_{D}(i)
{\mbox{\boldmath$\psi$}}_{D}(i)|^{2}, \label{eq:rankadaptationso}
\end{equation} where $\lambda_{D}$ is a forgetting factor.
Since the optimal rank can be considered as a function of the time
index $i$ \cite{MLHonig2002}, the forgetting factor is required and
allows us to track the optimal rank. We assume that the number of
branches $C$ and the length of the inner function $q$ are fixed. For
each time instant, we update a reduced-rank filter $\bar {\mathbf
w}_{\rm M}(i)$ and a projection vector
${\mbox{\boldmath$\psi$}}_{M}(i)$ with the maximum rank $D_{\rm
max}$, which can be expressed as
\begin{equation}
\begin{split}
& \bar {\mathbf w}_{\rm M}(i)=[\bar {w}_{\rm M,1}(i), \dots, \bar
{w}_{\rm M,\it D}(i), \dots, \bar{w}_ {\rm M,\it D_{\rm
max}}(i)]^{T}\\
&{\mbox{\boldmath$\psi$}}_{\rm M}(i)=[{\mbox{$\psi$}}_{\rm M,1}(i),
\dots, {\mbox{$\psi$}}_{\rm M,\it qD}(i), \dots,
{\mbox{$\psi$}}_{\rm M,\it qD_{\rm max}}(i)]^{T}.\\
\end{split}
\end{equation}
After the adaptation, we test values of $D$ within the range $D_{\rm
min}$ to $D_{\rm max}$. For each tested rank, we use the following
estimators
\begin{equation}
\begin{split}
& \bar {\mathbf w}_{D}(i)=[\bar {w}_{\rm M,1}(i), \dots,
\bar {w}_{\rm M,\it D}(i)]^{T}\\
&{\mbox{\boldmath$\psi$}}_{D}(i)=[{\mbox{$\psi$}}_{\rm
M,1}(i), \dots, {\mbox{$\psi$}}_{\rm M,\it qD}(i)]^{T}.\\
\end{split}
\label{eq:dadaptparameters}
\end{equation}
The position matrices for different model orders can be pre-stored
and the instantaneous position matrix $\mathbf P_{D}(i)$ can be
determined by the decision rule as shown in \eqref{eq:copt}. After
selecting the position matrix and given the input data matrix, we
substitute \eqref{eq:dadaptparameters} into
\eqref{eq:rankadaptationso} to obtain the value of
$\mathscr{C}_{D}(i)$, where $D\in\{D_{\rm min},\ldots,D_{\rm
max}\}$. The proposed algorithm can be expressed as
\begin{equation}
D_{\rm opt}(i)=\arg \min_{D\in\{D_{\rm min},\ldots,D_{\rm
max}\}}\mathscr{C}_{D}(i). \label{eq:rankadp}
\end{equation}
We remark that the complexity of updating the reduced-rank filter
and the projection vector in the proposed rank adaptation algorithm
is the same as the SAABF (C,$\rm D_{\rm max}$,q), since we only
adapt the $\bar {\mathbf w}_{\rm M}(i)$ and
${\mbox{\boldmath$\psi$}}_{\rm M}(i)$ for each time instant.
However, additional computations are required for calculating the
values of $\mathscr{C}_{D}(i)$ and selecting the minimum value of a
$(D_{\rm max}-D_{\rm min}+1)$-dimensional vector that corresponds to
a simple search and comparison.

\subsection{Inner Function Length Selection}

In the SAABF scheme, the length of the inner function is also a sensitive
parameter that affects the complexity and the overall performance. In this
work, we apply the similar idea used for the rank adaptation, to select the
optimal value of $q$. The criterion to choose $q_{\rm opt}$ is that it
minimizes the following cost function
\begin{equation}
\mathscr{C}_{q}(i)=\sum^{i}_{n=0} \lambda_{q}^{i-n}|d(i)-\bar
{\mathbf{w}}^{H}(i) \mathbf R_{\rm in}(i)\mathbf P_{q}(i)
{\mbox{\boldmath$\psi$}}_{q}(i)|^{2},
\end{equation} where the forgetting factor $\lambda_{q}$ is applied,
since we observe that in the SAABF scheme, the length of $q$ plays a
similar role as the rank $D$ and the optimal $q$ can change as a
function of the time index $i$.

When the model order $D$ and the branch number $C$ are fixed, for
each time instant, we adapt a $D$-by-$1$ reduced-rank filter $\bar
{\mathbf{w}}(i)$ jointly with a $Dq_{\rm max}$-by-$1$ projection
vector ${\mbox{\boldmath$\psi$}}_{\rm Q}(i)=[{\mbox{$\psi$}}_{\rm
Q,1}(i), \dots, {\mbox{$\psi$}}_{\rm Q,\it Dq}(i), \dots, {\mbox{$\psi$}}_{\rm Q,\it Dq_{\rm max}}(i)]^{T}.$
For different values of $q$, we use the estimate
\begin{equation}
{\mbox{\boldmath$\psi$}}_{q}(i)=[{\mbox{\boldmath$\psi$}}_{q,1}^{T}(i),
\dots, {\mbox{\boldmath$\psi$}}_{q,D}^{T}(i)]^{T},
\label{eq:Qadaptparameters}
\end{equation} where the vectors of ${\mbox{\boldmath$\psi$}}_{q,d}(i)$, $d=1, \dots, D$, can be expressed as
\begin{equation}
{\mbox{\boldmath$\psi$}}_{\rm q,d}(i)=[{\mbox{$\psi$}}_{\rm
Q,\it (d-1)q_{\rm max}+1}(i), \dots, {\mbox{$\psi$}}_{\rm Q,\it (d-1)q_{\rm max}+q}(i)]^{T}.\\
\end{equation}
At the $i$-th moment, we search from $q_{\rm min}$ to $q_{\rm max}$
and determine the $q_{\rm opt}$ using the following algorithm
\begin{equation}
q_{\rm opt}(i)=\arg \min_{q\in\{q_{\rm min}, \ldots, q_{\rm
max}\}}\mathscr{C}_{q}(i).\label{eq:qadaptationso}
\end{equation}
The computational complexity of updating the reduced-rank filter and
the projection vector in this algorithm is the same as the SAABF
(C,D,$\rm q_{max}$). Since we only adapt a $D$-by-$1$ reduced-rank
filter and a $Dq_{\rm max}$-by-$1$ projection vector for all tested
values of $q$. Additional computations are needed to compute the
values of $\mathscr{C}_{q}(i)$ and search the minimum value in a
$(q_{\rm max}-q_{\rm min}+1)$-dimensional vector.


\section{Simulations}
\label{sec:simulations}

In this section, we apply the proposed generic and SAABF schemes to
the uplink of a multiuser BPSK DS-UWB system and evaluate their
performance against existing reduced-rank and full-rank methods. 
In all numerical simulations, all the users are assumed
to be transmitting continuously at the same power level. The pulse
shape adopted is the RRC pulse with the pulse-width $0.375$ns. The
spreading codes are generated randomly for each user in each
independent simulation with a spreading gain of $32$ and the data
rate of the communication is approximately $83$Mbps. The standard
IEEE 802.15.4a channel model for the NLOS indoor environment is
employed \cite{Molisch2005} and we assume that the channel is
constant during the whole transmission. The channel delay spread is
$T_{DS}=30ns$ that is much larger than the symbol duration, which is
$T_{s}=12ns$. Hence, the severe ISI from $2G=6$ neighbor symbols are
taken into the account for the simulations. The sampling rate at the
receiver is assumed to be $2.67$GHz and the length of the discrete
time received signal is $M=112$. For all the simulations, the
adaptive filters are initialized as null vectors. This allows a fair
comparison between the analyzed techniques for their convergence
performance. In practice, the filters can be initialized with prior
knowledge about the spreading code or the channel to accelerate the
convergence. In this work, we present the uncoded bit error rate
(BER) for all the comparisons. All the curves are obtained by
averaging $200$ independent simulations.

\begin{figure}[htb]
\begin{minipage}[b]{1.0\linewidth}
  \centering
  \centerline{\epsfig{figure=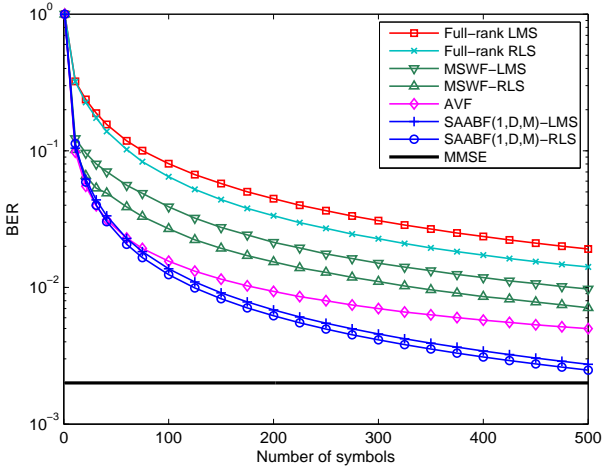,scale=0.6}}
\end{minipage}
\caption{BER performance of different algorithms for a SNR=20dB and 8 users.
The following parameters were used: full-rank LMS ($\mu=0.075$), full-rank RLS
($\lambda=0.998$, $\delta=10$), MSWF-LMS ($D=6$, $\mu=0.075$), MSWF-RLS ($D=6$,
$\lambda=0.998$), AVF ($D=6$), SAABF (1,3,M)-LMS ($\mu_{w}=0.15$,
$\mu_{\psi}=0.15$, 3 iterations) and SAABF (1,3,M)-RLS ($\lambda=0.998$,
$\delta=10$, 3 iterations).} \label{fig:jointvsother}
\end{figure}
\begin{figure}[htb]
\begin{minipage}[b]{1.0\linewidth}
  \centering
  \centerline{\epsfig{figure=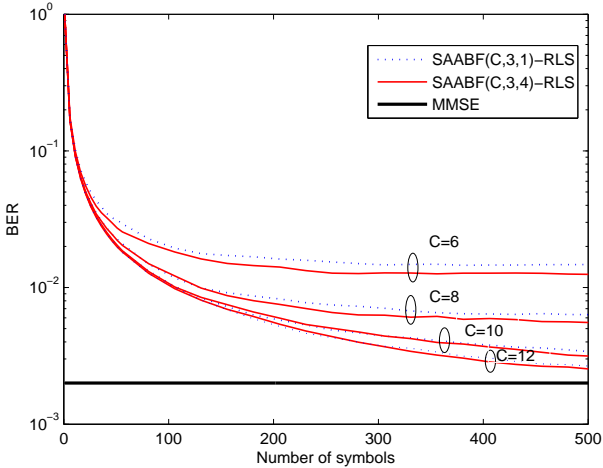,scale=0.6}}
\end{minipage}
\caption{BER performance of the proposed SAABF scheme versus the number of
training symbols for a SNR=20dB. The number of users is 8 and the following
parameters were used: SAABF-RLS ($\lambda=0.998$, $\delta=10$).}
\label{fig:saabfq1and4}
\end{figure}
\begin{figure}[htb]
\begin{minipage}[b]{1.0\linewidth}
  \centering
  \centerline{\epsfig{figure=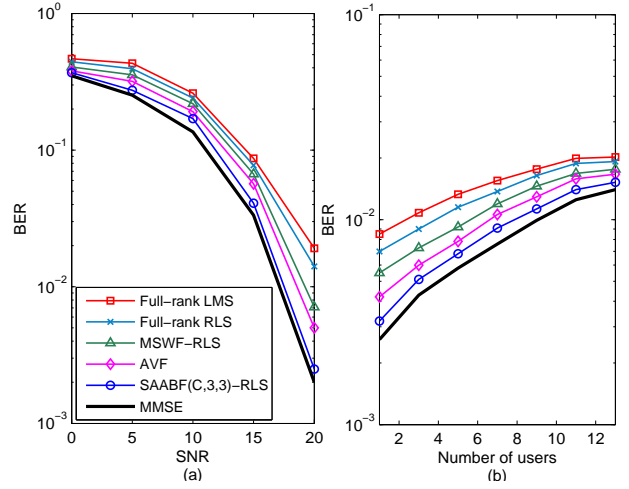,scale=0.6}}
\end{minipage}
\caption{BER performance of the proposed scheme with different SNRs and number
of users.} \label{fig:bervsusersnr}
\end{figure}
\begin{figure}[htb]
\begin{minipage}[b]{1.0\linewidth}
  \centering
  \centerline{\epsfig{figure=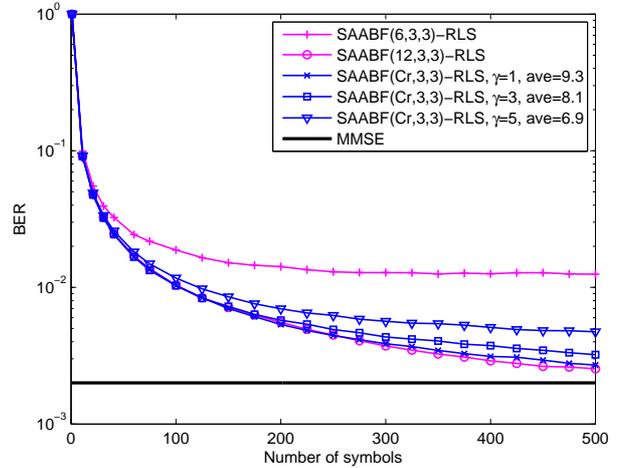,scale=0.6}}
\end{minipage}
\caption{BER performance of the SAABF scheme with branch-number selection. The
scenario of 20dB and 8 users are considered. The parameters used: SAABF-RLS
($\lambda=0.998$, $\delta=10$). For branch-number selection algorithm:  $C_{\rm
min}=6$ and $C_{\rm max}=12$, threshold $\gamma$ is in the unit of dB.}
\label{fig:cselection}
\end{figure}
\begin{figure}[htb]
\begin{minipage}[b]{1.0\linewidth}
  \centering
  \centerline{\epsfig{figure=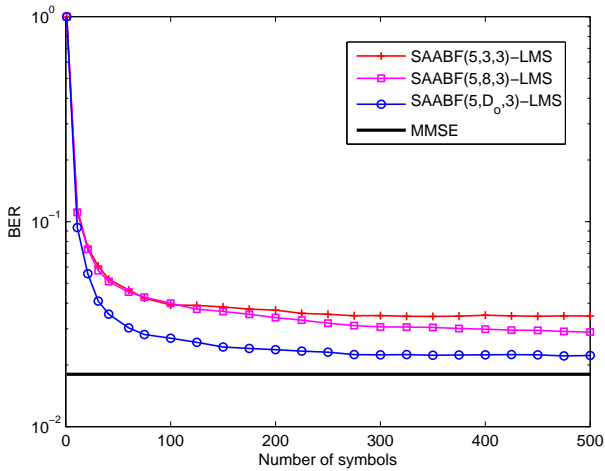,scale=0.6}}
\end{minipage}
\caption{BER performance of the SAABF scheme with rank adaptation. The scenario
of 16dB and 8 users are considered. The parameters used: SAABF-LMS
($\mu_{w}=0.15$, $\mu_{\psi}=0.15$). For rank-adaptation algorithm: $D_{\rm
min}=3$, $D_{\rm max}=8$ and $\lambda_{D}=0.998$.} \label{fig:dselection}
\end{figure}
\begin{figure}[htb]
\begin{minipage}[b]{1.0\linewidth}
  \centering
  \centerline{\epsfig{figure=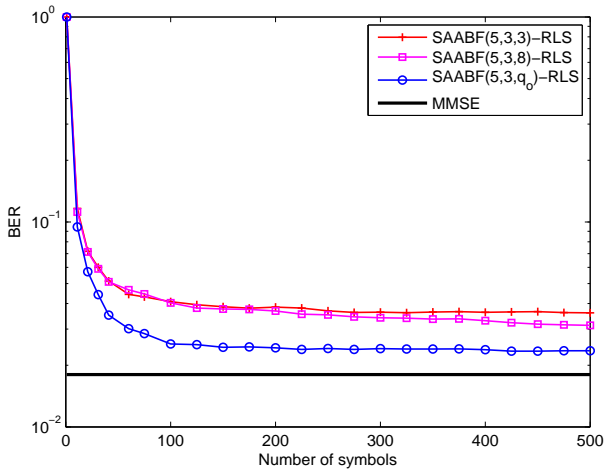,scale=0.6}}
\end{minipage}
\caption{BER performance of the SAABF scheme with adaptive short function
length. The scenario of 16dB and 8 users are considered. The parameters used:
SAABF-RLS ($\lambda=0.998$, $\delta=10$). $q_{\rm min}=3$, $q_{\rm max}=8$ and
$\lambda_{q}=0.998$.} \label{fig:qselection}
\end{figure}
\begin{figure}[htb]
\begin{minipage}[b]{1.0\linewidth}
  \centering
  \centerline{\epsfig{figure=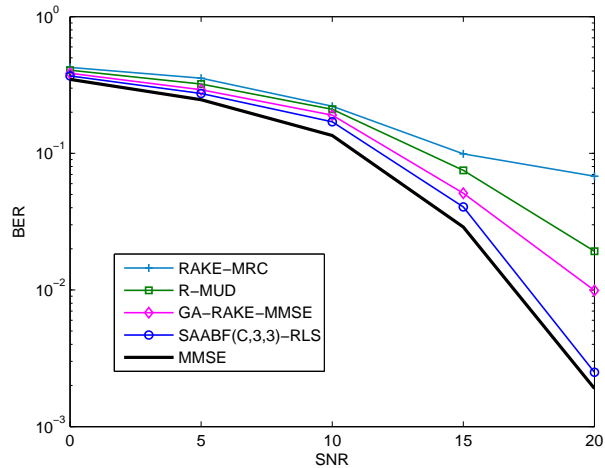,scale=0.6}}
\end{minipage}
\caption{BER performance against SNR of different receiver structures in a
system with 8 users.} \label{fig:diffreceivers}
\end{figure}

The first experiment we perform is to compare the uncoded BER performance of
the generic reduced-rank scheme, which is denoted as SAABF (1,D,M), with the
full-rank LMS and RLS algorithms, the LMS and RLS versions of the MSWF, and the
AVF method. We consider the scenario with a signal-to-noise ratio (SNR) of
$20$dB, $8$ users. 
Fig.3 shows the BER performance of different schemes as a function
of training symbols transmitted. The proposed generic scheme
outperforms all the other methods with 3 iterations. In the generic
scheme, the joint RLS algorithm could converge faster than the joint
LMS algorithm with the same number of iterations.

Fig.\ref{fig:saabfq1and4} shows the uncoded BER performance of the RLS version
of the novel SAABF scheme with different number of branches in the same
scenario as in the first experiment. In this experiment, the performance of the
simple configuration SAABF (C,D,1) is compared with SAABF (C,D,q), where $q=4$.
Note that, in SAABF (C,D,1), the projection vector ${\mbox{\boldmath$\psi$}}
(i)$ is no longer updated, we use its initial value for the whole transmission.
In the SAABF (C,D,q) scheme, when a sufficient number of branches are employed,
both versions of the joint adaptive algorithm can achieve excellent performance
with only one iteration for each input data. 
Increasing the number of branches, the performance
approaches that of the full-rank MMSE filter. The SAABF (C,D,1)
scheme can achieve a similar convergence speed to the SAABF (C,D,q),
but the SAABF (C,D,q) has better steady-state performance.

Fig.\ref{fig:bervsusersnr} (a) and (b) show the uncoded BER performances of
algorithms with different SNRs in a 8 users communication and with different
numbers of users in a 18dB scenario, respectively. It should be noted that if
the number of training symbols is sufficient, the performance of the full-rank
algorithms and the reduced-rank algorithms will approach the performance of the
full-rank MMSE filter. However, for short data support the reduced rank
algorithms outperform the full-rank algorithms due to their faster training. In
these experiments, 500 symbols are transmitted for each tested environment in
each independent simulation. The SAABF (C,3,3)-RLS is employed with C in the
range of 2 to 12. For different scenarios, the minimum number of branches that
enables the proposed scheme to approach the linear MMSE
performance is chosen. 
The novel SAABF scheme outperforms all other schemes in all the
simulated scenarios. 

The uncoded BER performance of the proposed RLS version of the SAABF scheme
with the implementation of the branch number selection algorithm is shown in
Fig.\ref{fig:cselection}. The proposed algorithm instantaneously chooses the
number of branches $C_{r}$ using \eqref{eq:cselectioneq}, from the range
$C_{\rm min}=6$ to $C_{\rm max}=12$. As the threshold $\gamma$ increasing, the
average required number of branches $C_{r}$ and the overall complexity are
reducing, but the performance degrading. For a $1$dB threshold, the performance
of the branch number selection SAABF ($C_{r}$,D,q) is very close to the SAABF
($C_{\rm max}$,D,q), while the average branch number $C_{r}$ is only $9.3$,
which is considerably lower than the $C_{\rm max}=12$. Hence, with the branch
number selection algorithm we obtain a solution which has lower complexity and
similar performance to that when the $C_{\rm max}$ is used.

Fig.\ref{fig:dselection} compares the BER performance of the SAABF-LMS using
the rank-adaptation algorithm with $C=5$ and $q=3$. The results using a
fixed-rank of $3$ and $8$ are also shown in Fig.\ref{fig:dselection} for
comparison purposes and illustration of the sensitivity of the SAABF scheme to
the rank $D$. The rank-adaptation solution selects the optimal rank $D_{o}(i)$
using \eqref{eq:rankadp} for each time instant, from the range $D_{\rm min}=3$
to $D_{\rm max}=8$. 
The BER performance of the SAABF scheme with the
rank-adaptation algorithm outperforms the fixed-rank SAABF scheme
with $D_{\rm min}$ or $D_{\rm max}$. In this environment, $D=8$ has
better steady-state performance than $D=3$, with both cases
showing the same convergence speed. 

Fig.\ref{fig:qselection} shows the BER behavior of the
SAABF-RLS scheme equipped the adaptive algorithm that determining
the length of the inner function with $C = 5$ and $D = 3$. The value
of $q_{o}(i)$ for each time instant is determined by
\eqref{eq:qadaptationso} with $q_{\rm min}=3$ and $q_{\rm max}=8$. A
clear improvement is shown when the algorithm that selects $q$ is
used.

In the last experiment, we conduct a comparison of the proposed and
existing linear receiver structures as shown in Fig. \ref{fig:diffreceivers}.
In a system with 8 users, we examine the performance of the traditional RAKE
receiver with the maximal-ratio combining (MRC), the reduced-order multiuser
detection (RMUD) \cite{yTian2006} with 15 taps, the generic algorithm (GA)
based RAKE-MMSE receiver \cite{SB2005} with 25 fingers and 20 iterations and
the proposed SAABF-RLS scheme (the parameters are the same as in
Fig.\ref{fig:bervsusersnr}). For each independent run, 500 symbols are
transmitted. The receiver with the SAABF-RLS scheme outperforms other receiver
structures especially in high SNR scenarios. Compared with the GA-RAKE-MMSE
scheme, a 2dB gain is obtained for a BER around $10^{-2}$. The proposed SAABF
scheme is able to suppress the interference efficiently without the knowledge
of the channel, the noise variance and the spreading codes.

\section{Conclusions}
\label{sec:conclusion}

In this work, we have introduced a generic reduced-rank scheme for interference
suppression, which jointly updates the projection vector and the reduced-rank
filter. Then, by constraining the design of the projection vector in the
generic scheme, we investigated a novel reduced-rank interference suppression
scheme based on switched approximations of adaptive basis function (SAABF) for
DS-UWB system. LMS and RLS algorithms were developed for adaptive estimation of
the parameters of the SAABF scheme. The uncoded BER performance of the novel
receiver structure was then evaluated in various scenarios with severe MAI and
ISI. With a low complexity, the SAABF scheme outperforms other reduced-rank
schemes and full-rank schemes. A discussion of the global optimality of the
reduced-rank filter was presented, and the relationships between the SAABF and
the generic scheme and the full-rank scheme were established.


\appendices

\section{Analysis of the Optimization Problem}
\label{app:optdiscussion}

In this Appendix, we discuss the optimization problem of the
proposed SAABF scheme. Specially, we consider the convergence of the
SAABF scheme via the computation of the Hessian matrix of the MSE
cost function which can be expressed as
\begin{equation}
\mathbf J_{\rm MSE}(\bar{\mathbf
w}(i),\mbox{\boldmath$\psi$}(i))=E[|d(i)-\bar{\mathbf
w}^{H}(i)\mathbf R_{\rm in}(i)\mathbf
P(i)\mbox{\boldmath$\psi$}(i)|^{2}]. \label{eq:definedcfa1}
\end{equation}
It is known that the convexity of the function can be verified if its Hessian
matrix is positive semi-definite. However, 
the SAABF scheme includes a discrete optimization of the position
matrix and a continuous adaptation of the reduced-rank filter and
the projection vector. For the position matrix selection problem, 
we constrain the design of the position matrix to a small number of
pre-stored matrices and switch between these matrices to choose the
instantaneous sub-optimum position matrix. 
This feature of the SAABF scheme suggests that the optimum values of
the three variables of the MSE cost function may be difficult to
obtain together, and that there are multiple solutions of the cost
function. The convexity is only verified when we consider one of the
continuously adapted variables whilst the others are kept fixed.
Firstly, let us compute the $D$-by-$D$ Hessian matrix for
\eqref{eq:definedcfa1} with respect to the reduced-rank filter:
\begin{equation}
\begin{split}
\mathbf H_{J,\bar{\mathbf w}}&= \frac{\partial  ^{2} \mathbf {J}_{\rm
MSE}}{\partial {\bar {\mathbf{w}}^{H}(i)\partial {\bar
{\mathbf{w}}(i)}}}\\
&=E[\mathbf R_{\rm in}(i)\mathbf
P(i)\mbox{\boldmath$\psi$}(i)\mbox{\boldmath$\psi$}^{H}(i)\mathbf
P^{H}(i)\mathbf R_{\rm in}^{H}(i)].
\end{split}
\end{equation}
For any $D$-dimensional non-zero vector $\mathbf a$, we discuss the
following scale term
\begin{equation}
\begin{split}
\mathbf a ^{H}\mathbf H_{J,\bar{\mathbf w}}\mathbf a & = E[\mathbf a ^{H}\mathbf
R_{\rm in}(i)\mathbf
P(i)\mbox{\boldmath$\psi$}(i)\mbox{\boldmath$\psi$}^{H}(i)\mathbf
P^{H}(i)\mathbf R_{\rm in}^{H}(i)\mathbf a]\\
&=E[\hat a(i) \hat a^{*}(i)]=E[|\hat
a(i)|^{2}], \label{eq:scaletermconvex}
\end{split}
\end{equation} where $\hat a(i)=\mathbf a ^{H}\mathbf
R_{\rm in}(i)\mathbf P(i)\mbox{\boldmath$\psi$}(i)$. Assume that the
position matrix $\mathbf P(i)$ and the projection vector
$\mbox{\boldmath$\psi$}(i)$ are fixed. The scale term in
\eqref{eq:scaletermconvex} is always nonnegative. Hence, the Hessian
matrix $\mathbf H_{J,\bar{\mathbf w}}$ is a positive semi-definite
matrix. Similarly, the $qD$-by-$qD$ Hessian matrix for
\eqref{eq:definedcfa1} with respect to the projection vector is
\begin{equation}
\mathbf H_{J,\mbox{\boldmath$\psi$}}=E[\mathbf P^{H}(i)\mathbf
R_{\rm in}^{H}(i)\bar{\mathbf w}(i)\bar{\mathbf w}^{H}(i)\mathbf
R_{\rm in}(i) \mathbf P(i)],
\end{equation} which is also a positive semi-definite matrix if the position matrix and the
reduced-rank filter are fixed.

In the SAABF scheme, after determined the position
matrix, the optimization problems for the projection vector and the
reduced-rank filter can be consider as a bi-convex problem
\cite{biconvex}: by fixing one of the parameters, the other design
problem is convex. In order to test the convergence of the SAABF
scheme in the case of jointly updating $\bar{\mathbf w}(i)$ and
$\mbox{\boldmath$\psi$}(i)$, we checked the impact of different
initializations, which confirmed that the performance of the
algorithms are not subject to degradation due to the initialization.
However, the proof of the global optimum and no local minima with
the joint adaptive algorithm remains an interesting open problem to
be researched.

\section{Proof of the Equivalence of the Schemes}
\label{app:equivalence}

In this section, we prove that the SAABF (1,D,M) is equivalent to the the
generic scheme and the SAABF (1,1,M) is equivalent to the full-rank scheme.

Firstly, we express the MMSE solutions for the SAABF scheme as
\begin{equation}
\bar {\mathbf{w}}_{\rm MMSE}=\bar{\mathbf R}^{-1}\bar{\mathbf
p},~{\mbox{\boldmath$\psi$}}_{\rm MMSE}={\mathbf R}_{\psi}^{-1}{\mathbf
p}_{\psi} \label{eq:MMSEsolutionfor-saabf}
\end{equation} where $\bar{\mathbf R}=E[\mathbf R_{\rm in}(i)\mathbf P(i)
{\mbox{\boldmath$\psi$}}(i) {\mbox{\boldmath$\psi$}}^{H}(i)\mathbf
P^{H}(i)\mathbf R_{\rm in}^{H}(i)]$, $ \bar{\mathbf
p}=E[d^{*}(i)\mathbf R_{\rm in}(i)\mathbf P(i)
{\mbox{\boldmath$\psi$}}(i)]$, ${\mathbf R}_{\psi}=E[\mathbf
P^{H}(i)\mathbf R_{\rm in}^{H}(i)\bar{\mathbf w}(i+1)\bar{\mathbf
w}^{H}(i+1)\mathbf R_{\rm in}(i)\mathbf P(i)]$ and ${\mathbf
p}_{\psi}=E[d(i)\mathbf P^{H}(i)\mathbf R_{\rm
in}^{H}(i)\bar{\mathbf w}(i+1)]$.
Revisit the expression of the basis functions in the SAABF scheme in
\eqref{eq:forzcd}. In SAABF (1,D,M), the length of the inner function equals to
the length of the basis function and the position matrix in \eqref{eq:t_c}
becomes an $MD$-by-$MD$ identity matrix. Hence, the MMSE solutions of the
generic scheme shown in \eqref{eq:mmsesolution-generic} are the same as
\eqref{eq:MMSEsolutionfor-saabf} when $\mathbf P(i)$ is an identity matrix,
which means the SAABF (1,D,M) is equivalent to the generic scheme.

Secondly, we prove that the SAABF (1,1,M), or the generic scheme with $D$=$1$,
is equivalent to the full-rank scheme in the sense that they have the same MMSE
corresponding to the optimum solutions. 
Here, we expand the cost function of the generic scheme that is
shown in \eqref{eq:costfuntionfullranknew}
\begin{equation}
\begin{split}
\mathbf J_{\rm G} &=\sigma^{2}_{d}-E[d(i)\mathbf t^{H}(i)\mathbf R_{\rm
in}^{H}(i)\bar{\mathbf w}(i)]\\
&-E[d^{*}(i)\bar{\mathbf w}^{H}(i)\mathbf R_{\rm
in}(i)\mathbf t (i)]\\
&+E[\bar{\mathbf w}^{H}(i)\mathbf R_{\rm in}(i)\mathbf
t(i)\mathbf t^{H}(i)\mathbf R_{\rm in}^{H}(i)\bar{\mathbf w}(i)].
\end{split}
\label{eq:J-g}
\end{equation}
In the case of $D$=$1$, the input data matrix $\mathbf R_{\rm in}(i)$=$\mathbf
r^{T}(i)$ becomes a $1$-by-$M$ vector, the reduced-rank filter has only one tap
and hence the $\bar{w}_{\rm opt}$ is a scalar term and we can find the
relationship between $\mathbf t_{\rm opt}$ and $\bar{w}_{\rm opt}$ as
\begin{equation*}
\begin{split}
\mathbf t_{\rm opt}&=(E[\mathbf r^{*}(i)\mathbf r^{T}(i)] \bar{w}_{\rm opt}
\bar{w}_{\rm opt}^{*})^{-1}E[d(i)\mathbf r^{*}(i)]\bar{w}_{\rm opt}\\
&=(\mathbf {R}^{T})^{-1}\mathbf p^{*} (\bar{w}_{\rm opt}^{*})^{-1}
\end{split}
\end{equation*}
Hence, the second term in \eqref{eq:J-g} becomes
\begin{equation}
\begin{split}
&E[d(i)\mathbf t^{H}(i)\mathbf R_{\rm in}^{H}(i)\bar{\mathbf w}(i)]=\mathbf
t_{\rm opt}^{H} E[d(i)\mathbf r^{*}(i)] \bar{w}_{\rm opt}\\
&=[(\mathbf {R}^{T})^{-1}\mathbf p^{*}(\bar{w}_{\rm opt}^{*})^{-1}]^{H}\mathbf p^{*}\bar{w}_{\rm opt}
=\mathbf p^{T}(\mathbf {R}^{T})^{-1}\mathbf p^{*}\\
&=(\mathbf p^{H}\mathbf R^{-1}\mathbf p)^{T}=\mathbf p^{H}\mathbf R^{-1}\mathbf p.\\
\end{split}
\label{eq:J-gt2}
\end{equation} Note that here we use the fact that the transpose of the scale term
$\mathbf p^{H}\mathbf R^{-1}\mathbf p$ is itself and $(\mathbf
{R}^{T})^{H}=\mathbf {R}^{T}$. Since the third scalar term in \eqref{eq:J-g} is
the conjugate of the second term, we have $E[d^{*}(i)\bar{\mathbf
w}^{H}(i)\mathbf R_{\rm in}(i)\mathbf t (i)]=(\mathbf p^{H}\mathbf
R^{-1}\mathbf p)^{H}=\mathbf p^{H}\mathbf R^{-1}\mathbf p.$
The fourth term of \eqref{eq:J-g} can be expanded as
\begin{equation}
\begin{split}
&E[\bar{\mathbf w}^{H}(i)\mathbf R_{\rm in}(i)\mathbf t(i)\mathbf t^{H}(i)\mathbf R_{\rm in}^{H}(i)\bar{\mathbf w}(i)]\\
&=\bar{w}_{\rm opt}^{*} E[\mathbf r^{T}(i)\mathbf t_{\rm opt}\mathbf t^{H}_{\rm opt}\mathbf r^{*}(i)]\bar{w}_{\rm opt}\\
&=\bar{w}_{\rm opt}^{*} E[\mathbf t^{H}_{\rm opt}\mathbf r^{*}(i)\mathbf
r^{T}(i)\mathbf t_{\rm opt}]\bar{w}_{\rm opt}\\
&=\mathbf p^{T}(\mathbf {R}^{T})^{-1}\mathbf
p^{*}=\mathbf p^{H}\mathbf R^{-1}\mathbf p.
\end{split}
\label{eq:J-gt4}
\end{equation}
Hence, the MMSE of the generic scheme for $D$=$1$ is $\mathbf J_{\rm
GMMSE}=\sigma^{2}_{d}-\mathbf p^{H}\mathbf R^{-1}\mathbf p,$ which is the same
as the MMSE obtained via the full-rank Wiener filter as shown in
\eqref{eq:MMSE-full-rank}. This completes the proof.

\end{document}